# Phase relations in boron at pressures up to 18 GPa and temperatures up to 2200 °C

Jiaqian Qin[1*], Tetsuo Irifune[1], Haruhiko Dekura[2], Hiroaki Ohfuji[1], Norimasa Nishiyama[1], Li Lei[1], Toru Shinmei[1]

[1]Geodynamics Research Center, Ehime University, 2-5 Bunkyo-cho, Matsuyama, 790-8577, Japan

[2]Senior Research Fellow Center, Ehime University, 2-5 Bunkyo-cho, Matsuyama, 790-8577, Japan

The phase relations in boron have been investigated at high pressure and high temperature using a multianvil apparatus, and the quenched sample has been analyzed by x-ray diffraction, Raman spectra and transmission electron microscopy. We demonstrate that $\gamma$-$B_{28}$ can be synthesized over a wide P-T range, and T-$B_{50}$ is obtained at higher temperatures and similar pressures. The phase boundary of the $\beta$-$B_{106}$, $\gamma$-$B_{28}$ and T-$B_{50}$ is determined at pressures between 7 and 18 GPa and the temperatures of 500-2200 °C. The results suggest that T-$B_{50}$ might be an intermediate phase-metastable, formed for kinetic reasons (Ostwald rule) on the way from $\beta$-$B_{106}$ to T-192 and $\gamma$-$B_{28}$ to T-192.

PACS number(s): 64.60.-i

[*] Corresponding author, E-mail: jiaqianqin@gmail.com

Boron has been widely studied due to its complex polymorphism and fascinating chemical and physical properties.[1-12] All known structures of boron and some boron-rich compounds contain $B_{12}$ icosahedron, which can be flexibly linked into rigid framework structures. Many previous studies of boron concentrated on the synthesis and structural characterization of pure forms. Now probably four of the reported boron phases correspond to the pure element.[1,2,5,13-16] : α-rhombohedral $B_{12}$ (α-$B_{12}$), β-rhombohedral $B_{106}$ (β-$B_{106}$), β-tetragonal $B_{192}$ (T-192), and γ-$B_{28}$, but α-tetragonal $B_{50}$ (T-$B_{50}$ or T-$B_{52}$) is also believed to exist[17]. In particular, γ-$B_{28}$ turned out to be the key to understanding the phase diagram of boron-boron is the only element for which the phase diagram was unknown since its discovery 200 years ago. In previous studies, Oganov *et al.*[1] gave a phase diagram of boron based on their work and previous experiments[5,8,18] and theoretical studies[19]. Zarechnaya *et al.*[2] synthesized the γ-$B_{28}$ from β-$B_{106}$ and gave tentative phase relations between γ-$B_{28}$ and β-$B_{106}$ based on two multianvil quenched experiments and *in situ* DAC experiments. Shirai *et al.*[7,20,21] studied the properties of β-$B_{106}$ and the phase diagram of β-$B_{106}$ and α-$B_{12}$ by *ab initio* calculations. T-192 phase was discovered at Polytechnic Institute of Brooklyn in 1960[14] (the structure of the latter was so complex that it was solved only in 1979[15]). Amberger and Ploog observed amorphous boron, α-$B_{12}$ and T-192 phases, and occasionally β-$B_{106}$[16]. Ma *et al.*[5] and Oganov *et al.*[1] found that the formation of the T-192 phase at HPHT using a DAC and multianvil press, respectively. The single crystal T-$B_{50}$ crystal was obtained by Laubengayer *et al.*[22] The crystal structure of T-$B_{50}$ was proposed by Hoard *et al.* (P42/nnm (134), a=8.75 Å and c=5.06 Å)[23,24]. However, no other experimental results of T-$B_{50}$ was reported so far, so it was believed to be hypothetical boron phase, which can exists only as $B_{50}N_2$, $B_{50}C_2$ or some other compounds [25,26]. The main reason may mainly be ascribed to the complexity of the boron crystal structures themselves. The complex crystal structure also cause that the relative stability of boron phases is still experimentally unresolved even at ambient conditions.

In this Rapid Communication, we conducted a series of experiments on the HPHT synthesis of γ-$B_{28}$ and T-$B_{50}$ from β-$B_{106}$ (claimed purity 99.6%, Goodfellow) and examined β-$B_{106}$ at HPHT using a multianvil press (see Ref [27], Supplementary materials), and quenched the sample to ambient conditions for XRD and Raman spectra to investigate the phase relations of β-$B_{106}$, γ-$B_{28}$ and T-$B_{50}$. Table S1 (see Ref [27]) shows the synthesis experimental conditions and the characterization of the recovered samples. The T-$B_{50}$ samples were also loaded into a focused ion beam (FIB) system (JOEL JEM-9310FIB) to prepare transmission electron microscopy (TEM)-foils. The detailed procedure of FIB milling for TEM-foil preparation is described elsewhere[28]. TEM observations were carried out using a JEOL JEM-2010, operated at 200 kV. Selected area electron diffraction (SAED) was employed to characterize T-$B_{50}$.

Fig. 1 shows XRD patterns and Raman spectra of the recovered products that were heated 1600 ºC for 2 hours at 15.0 GPa and the starting materials. From the XRD patterns [Fig. 1(a)], we find that $γ$-$B_{28}$ is synthesized under this high P-T condition. Besides the $γ$-$B_{28}$, further phase identification shows that the HPHT quenched sample also contain small amounts of MgO (MgO capsule). In addition, the XRD data of $γ$-boron from Oganov *et al.*[1] are also shown in Fig. 1(a), and we can find that our XRD patterns are in good agreement with the reported results. The recovered samples were also investigated using Raman spectra. The Raman spectra of the recovered sample are also different from $β$-$B_{106}$ [Fig. 1(b)], but are similar to that of $γ$-$B_{28}$ reported by Zarechnaya *et al.*[2,29]. Therefore, we confirm that $γ$-$B_{28}$ is successfully synthesized under this high P-T condition. In addition, we also tried to synthesize $γ$-$B_{28}$ under different high P-T conditions (Table S1). XRD and Raman spectra results indicate that $γ$-$B_{28}$ can also be synthesized at 13 GPa and high temperatures of 1300 ºC and 1600 ºC for 2 h, 10 min, and even for just 1 min, 15 GPa and 1800 ºC for 10 min, and 18 GPa and 2000 ºC for 10 min. The results demonstrate that the $γ$-$B_{28}$ can be synthesized over a wide P-T range. By indexing the powder x-ray diffraction data using Powder X[30], the structure of quenched samples was solved to be an orthorhombic unit cell ($a$=5.0570(6) Å, $b$=5.6165(7) Å, $c$=6.9815(9) Å, $V$=198.29(6) Å$^3$). Our results are comparable with previous experimental ($a$=5.0544 Å, $b$=5.6199 Å, $c$=6.9873 Å and $a$=5.0576 Å, $b$=5.6245 Å, $c$=6.9884 Å) and theoretical results[1,2,4,31,32].

T-$B_{50}$ is also based on $B_{12}$ icosahedra, Kurakevych *et al.* (unpublished, arXiv:1110.1731) observed the crystallization of tetragonal boron T-$B_{50}$ from amorphous boron at HPHT. In our experiment, we studied $β$-$B_{106}$ at 2000 °C and high pressures of 9.5 and 12 GPa, and the quenched sample is pure boron using EDS analysis (see Ref [27], Figure S1). Fig. 2(a) shows the Raman spectra of the recovered products that were heated 2000 ºC for 5 min at 12.0 GPa and $β$-$B_{106}$. Because boron is a light element and poor x-ray scatterer, we did not obtain a good quality of XRD patterns. However, Raman spectra is easy to clarify the phase transition of boron, we can find different color zone from the optical image [Fig. 2(b)], and Raman spectra shows that the light gray and gray zone correspond to $\gamma$-B$_{28}$, but Raman spectra of dark gray are different from the $\beta$-B$_{106}$, $\alpha$-B$_{12}$[4] and $\gamma$-B$_{28}$, which were obtained the same results can be found at 15 GPa, 2100 ºC, and 18 GPa, 2200 ºC. In addition, we also conducted $\beta$-B$_{106}$ at 9.5 GPa and similar temperature, and quenched sample to ambient conditions for Raman spectra. Some special grains appear in the sample from the microscopic image [Fig. 2(c)], and Raman spectra of these grains are same as the dark gray of the quenched sample under higher pressure and other area are $\beta$-B$_{106}$. As the XRD quality is very low, and it is difficult to identify phase, we prepared TEM-foils (dark gray zone in Fig. 2(b)) using FIB system for TEM observations. Fig. 2(d,e) shows the TEM image and SAED patterns of TEM-foils. The SAED patterns can be indexed as the diffraction patterns of the [1-10] zone axis of T-B$_{50}$. And the lattice parameters are a=8.71 Å and c=5.00 Å, which is in good agreement with the reference (a=8.73 Å, c=5.03 Å[22] and a=8.75 Å, c=5.06 Å [23,24]). Thus, we have observed the T-B$_{50}$ from $\beta$-B$_{106}$ at HPHT, and presented the Raman spectra of T-B$_{50.}$ Our results also suggest the T-B$_{50}$ is a pure boron phase and can be synthesized at HPHT.

In order to understand the relative phase stability of boron, we investigated the phase relations of $\gamma$-B$_{28}$, $\beta$-B$_{106}$ and T-B$_{50}$. Table S1 shows the synthesis experimental conditions and the characterization of the recovered samples, and Fig. 3 shows selected XRD patterns and Raman spectra of the quenched samples from HPHT conditions. Characteristic x-ray and Raman spectra of the starting material $\beta$-B$_{106}$ are also presented in the bottom curves (*a* in Fig. 3). At high temperatures of 1300 ºC, it has a similar XRD pattern [curve *b*, Fig. 3(a)] to that of $\beta$-B$_{106}$ and no phase transitions are detected at 8.5 GPa. At 9 GPa, we find that the peaks of $\gamma$-B$_{28}$ appear [curve *c*, Fig. 3(a)]. The peaks of $\beta$-B$_{106}$ decrease with increasing in pressure. When pressure increases up to 10 GPa, the diffraction pattern shows $\gamma$-boron and the peaks of $\beta$-B$_{106}$ almost disappear [curve *d*, Fig. 3(a)]. Raman spectra also demonstrate a drastic change. The Raman spectra of starting materials and quenched sample from 8.5 GPa and 1300 ºC [curves *a* and *b,* Fig. 3(b)] are dominated by the $\beta$-B$_{106}$ phonon modes, while the spectra of 9.0 GPa or higher [curves *c, d,* and *e,* Fig. 3(b)] significantly differ from the previous two. Based on the XRD results and Raman spectra, we confirm the phase transition from $\beta$-B$_{106}$ to $\gamma$-B$_{28}$ between 8.5 GPa and 9 GPa under 1300 ºC.

Samples that treated temperatures of 1000 °C and 1600 °C at different pressures were also investigated using XRD and Raman spectra. The samples usually underwent HPHT conditions for 2 h, although some samples were treated for 4 h or above 20 h under P-T conditions near the phase boundary and at relatively lower temperature. Combination of the XRD and Raman spectra results [see Ref [27], Table S1 and Fig. S2(a), Fig. S2(b), Fig. S2(c), Fig. S2(d)] reveals that the P-T conditions of the $\beta$-$B_{106}$ to $\gamma$-$B_{28}$ phase transition lie between 8.5 and 9 GPa, 9 and 9.5 GPa, under high temperatures of 1000 and 1600 °C, respectively.

To further understand the phase boundary of $\beta$-$B_{106}$ to T-$B_{50}$ and $\gamma$-$B_{28}$ to T-$B_{50}$, we conducted $\beta$-$B_{106}$ at pressures between 7 and 18 GPa and the temperature of 1700-2200 °C, and quenched the sample to ambient conditions for XRD and Raman spectra (Table S1). Under high pressures of 12, 15, and 18 GPa, when $\beta$-$B_{106}$ is heated up to 1800, 1900, and 2000 °C, respectively, XRD and Raman spectra indicate that pure $\gamma$-$B_{28}$ is successfully synthesized. At similar pressures, the T-$B_{50}$ is produced at higher temperatures of 2000-2200 °C (Table S1). However, it is difficult to synthesize the pure T-$B_{50}$ according to our experimental results. Under lower pressures of 7 and 8 GPa and high temperature of 2000 °C, the quenched sample is pure $\beta$-$B_{106}$ from XRD and Raman spectra. When we increase the pressure to 9 GPa, the quenched sample is coexistence of $\beta$-$B_{106}$ and T-$B_{50}$ at high temperature of 2000 °C, but it is pure $\beta$-$B_{106}$ at 1800 °C. Thus, we give a tentative phase boundary of $\gamma$-$B_{28}$ to T-$B_{50}$ and $\beta$-$B_{106}$ to T-$B_{50}$ according to our experimental results (Fig.4).

Fig.4 shows the P-T region of the phase relations of $\beta$-$B_{106}$, $\gamma$-$B_{28}$ and T-$B_{50}$, and a comparison with the results of Oganov et al.[1] and Zarechnaya et al.[2], and the inset shows the P-T conditions for synthesis of T-192 phase from Oganov et al.[1] and Ma et al.[5]. In this P-T region, our phase boundary of $\beta$-$B_{106}$ to $\gamma$-$B_{28}$ is in good agreement with Oganov's theoretical results and Zarechnaya's DAC result. And phase boundary of $\gamma$-$B_{28}$ to T-$B_{50}$ and $\beta$-$B_{106}$ to T-$B_{50}$ is similar with the phase boundary of $\gamma$-$B_{28}$ to T-192 phase and $\beta$-$B_{106}$ to T-192 phase from Oganov et al.[1] results. Our results indicate that the phase relations of $\beta$-$B_{106}$ to $\gamma$-$B_{28}$ has almost no pressure dependence, and firstly give the phase relations of $\gamma$-$B_{28}$ to T-$B_{50}$ and $\beta$-$B_{106}$ to T-$B_{50}$, and prove that T-$B_{50}$ phase is not only pure boron phase, but also has a stability field at high pressures and temperatures.

From the P-T relations of $\beta$-$B_{106}$, $\gamma$-$B_{28}$, and T-$B_{50}$, the quenched samples are $\beta$-$B_{106}$ at different temperatures and lower pressures of 7, 8, and 8.5 GPa. From 9 to 10 GPa, only $\beta$-$B_{106}$ is observed at temperatures below 900 °C, but we observe a formation of $\gamma$-$B_{28}$ in the both quenched samples at high temperatures. However, when the temperature increases up to 2000 °C, some little T-$B_{50}$ phase grains are found in the quenched sample, but the main phase is still $\beta$-$B_{106}$. From 12 to 18 GPa, the pure $\gamma$-$B_{28}$ is successfully synthesized at high temperatures (above 1300 °C), and parts of $\gamma$-$B_{28}$ will transform to T-$B_{50}$ at higher temperatures. Extrapolation of phase boundary of $\beta$-$B_{106}$ and $\gamma$-$B_{28}$, we find $\beta$+ T-$B_{50}$ is in the "phase stability" of $\beta$-$B_{106}$, and $\gamma$+ T-$B_{50}$ is in the "phase stability" of $\gamma$-$B_{28}$. This is why we synthesize mixture of $\beta$-$B_{106}$ and T-$B_{50}$ at lower pressure and mixture of $\gamma$-$B_{28}$ and T-$B_{50}$ at higher pressure. Additionally, we also examined the $\gamma$-$B_{28}$, mixture of $\gamma$-$B_{28}$ and T-$B_{50}$, and mixture of $\beta$-$B_{106}$ and T-$B_{50}$ at 7 GPa and 1700 °C, 12 GPa and 1600 °C, and 7 GPa and 1700 °C, respectively. XRD and Raman spectra results show that $\gamma$-$B_{28}$ and T-$B_{50}$ transforms back to $\beta$-$B_{106}$, and T-$B_{50}$ transforms back to $\gamma$-$B_{28}$ under the stability field of $\beta$-$B_{106}$ and $\gamma$-$B_{28}$, respectively. Combined with the previous results[1, 5] and our study, we find that $\beta$-$B_{106}$ transforms into $\gamma$-$B_{28}$ at high pressure and a certain temperatures, and then transforms back to $\beta$-$B_{106}$ at relative low pressures (~9 GPa), with increasing temperature up to 2000 °C, $\beta$-$B_{106}$ will transform to T-$B_{50}$. However, $\beta$-$B_{106}$ transforms into $\gamma$-$B_{28}$ and then transforms into T-$B_{50}$ at relative higher pressures (above ~10 GPa). These results show that $\gamma$-$B_{28}$ is stable phase under a certain pressures (above ~9 GPa) in good agreement with previous theoretical results[1]. However, we didn't observe the T-192 phase according to our experiment. Interestingly, our phase stability of T-$B_{50}$ is in the theoretical phase stability of T-192 phase, so we suggest that T-$B_{50}$ could exist as an intermediate phase-metastable along the phase boundary of $\gamma$-$B_{28}$ to T-192 phase and $\beta$-$B_{106}$ to T-192 phase from Oganov *et al.* [1] results. These results may be the reason of Ostwald rule of stages, which shows the possibility phases with intermediate values of free Gibbs energy, before the system has transformed into the thermodynamically stable phase with the most stable crystal structure. According to first principles calculations[1], the free enthalpy of T-$B_{50}$ should be much higher than T-192 at 0 K, while our experiments reveal that T-$B_{50}$ should be along the phase boundary of $\gamma$-$B_{28}$ to T-192 phase and $\beta$-$B_{106}$ to T-192 phase. The different could be explained by the non-negligible 0 K configurationally entropy of T-$B_{50}$, which strongly decreases the free enthalpy at high temperatures. This also allows us to reasonably explain why we didn't observe the T-192 phase. It shows that synthesis of T-192 should need much more higher temperature at similar pressure. Thus, more experimental and theoretical studies on structural and phase stability of boron are strongly recommended, especially the thermodynamic phase relation between T-$B_{192}$ and T-$B_{50}$ at HPHT. Here we also give the kinetic boundary of $β$-$B_{106}$ to $γ$-$B_{28}$. The results indicate that the kinetic barriers decrease with increasing pressure, and no $α$-$B_{12}$ was found in our high P-T conditions.

In summary, we have synthesized $γ$-$B_{28}$ and T-$B_{50}$ from $β$-$B_{106}$ under HPHT using a multianvil press, and investigated the phase relations of $β$-$B_{106}$, $γ$-$B_{28}$ and T-$B_{50}$. We demonstrate that $γ$-$B_{28}$ can be synthesized over a wide P-T range, and T-$B_{50}$ is obtained at higher temperatures and similar pressures, and suggest that T-$B_{50}$ might be an intermediate phase-metastable, formed for kinetic reasons (Ostwald rule) on the way from $β$-$B_{106}$ to T-192 and $γ$-$B_{28}$ to T-192.

Based on the XRD, Raman spectra and TEM results, we present the experimental phase relations in boron at pressures up to 18 GPa and temperatures up to 2200 °C, and suggest that $β$-$B_{106}$ and $γ$-$B_{28}$ will directly transform to T-$B_{50}$ at lower and higher pressure, respectively. Additionally, T-$B_{50}$ will transform back to $β$-$B_{106}$ and $γ$-$B_{28}$ during quenching under lower pressures and higher pressures, respectively. The results provide fundamental information for better understanding of the phase diagram of this interesting material. However, more experimental and theoretical studies on structural and mechanical properties are strongly recommended.

ACKNOWLEDGMENTS

This work is supported by JSPS Grants-in Aid for Scientific Research under Grant No. 22 00029 and G-COE Program. One of authors (J. Qin) would like to thank Leiming Fang, Fulong Wang and Yongtao Zou for helping in experiment, and thank Profs. A. R. Oganov and V. L. Solozhenko for valuable discussions.

249 **Figure Caption**

250 FIG. 1. (color online). XRD patterns and Raman spectra of $\beta$-$B_{106}$ treated at 15 GPa and 1600 °C
251 and $\beta$-$B_{106}$, and compared with previous experimental results from Oganov et al.[1] and
252 Zarechnaya et al.[2], (a) XRD patterns, (b) Raman spectra.

253 FIG. 2. (color online). (a) Raman spectra of T-$B_{50}$ synthesized at 2000 °C and high pressures of
254 9.5 and 12 GPa, and in comparison with Raman spectra of $\beta$-$B_{106}$ and $\gamma$-$B_{28}$. (b) Optical photo of
255 quenched sample at 12 GPa, (c) Optical photo of quenched sample at 9.5 GPa, (d) TEM image of
256 the single crystal foil, (e) SAED pattern shows the diffraction patterns of the [1-10] zone axis of
257 T-$B_{50}$.

258 FIG. 3. (color online). XRD patterns and Raman spectra of $\beta$-boron treated at HPHT and $\beta$-$B_{106}$.
259 (a) XRD patterns, (b) Raman spectra.

260 FIG. 4. (color online). Phase relations between $\beta$-$B_{106}$ (open circle), $\gamma$-$B_{28}$ (solid circle) and T-$B_{50}$
261 (inverse triangles). The semi-solid circles represent $\beta$-$B_{106}$ and $\gamma$-$B_{28}$ in coexistence, the open and
262 solid inverse triangles are $\beta$-$B_{106}$ and T-$B_{50}$ in coexistence and $\gamma$-$B_{28}$ and T-$B_{50}$ mixture,
263 respectively, and the triangles represent P-T conditions of T-192 phase from Oganov et al.[1] and
264 Ma et al.[5], respectively. The line is a phase boundary of $\beta$-$B_{106}$, $\gamma$-$B_{28}$ and T-192 boron, and the
265 inset show the theoretical phase boundary from Oganov et al.[1] and the tentative phase boundary
266 from Zarechnaya et al.[2]. Additionally, reverse experiments are also shown in this figure, square
267 is phase transition from $\gamma$-$B_{28}$ to $\beta$-$B_{106}$, open and solid stars present T-$B_{50}$ transforms back to $\beta$-
268 $B_{106}$ and $\gamma$-$B_{28}$, respectively.

269

270

271

272

273

274

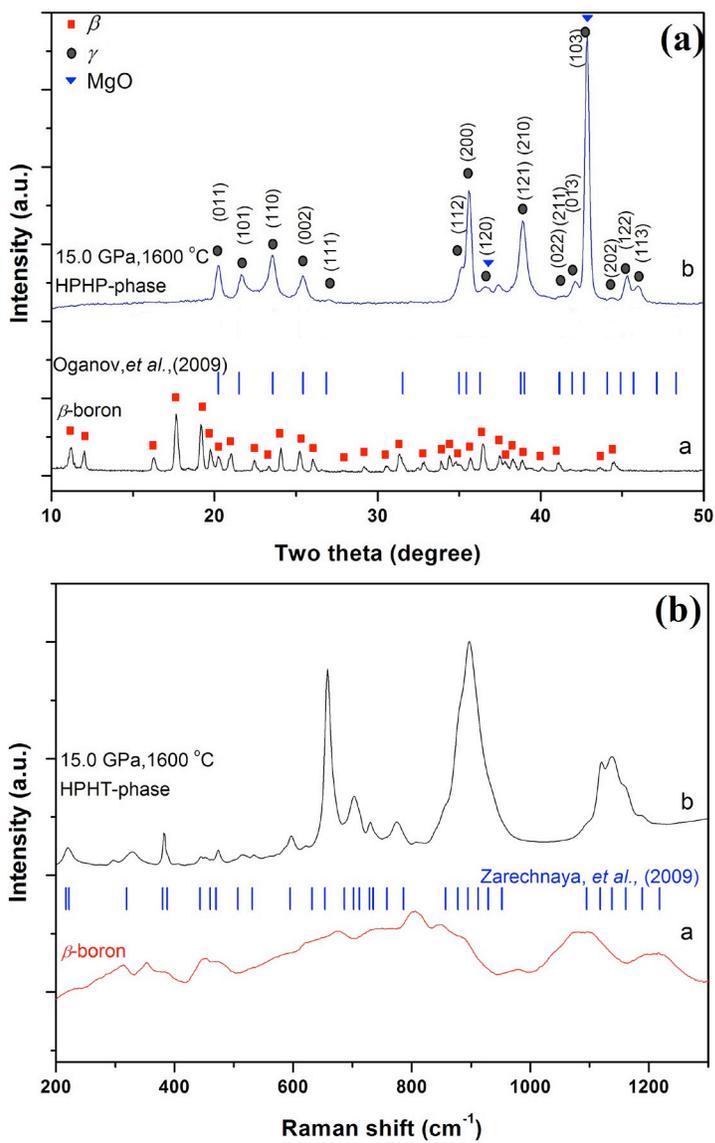

Fig.1

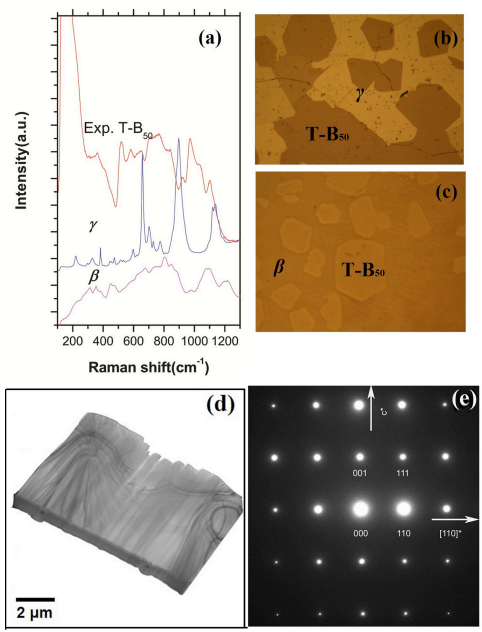

Fig.2

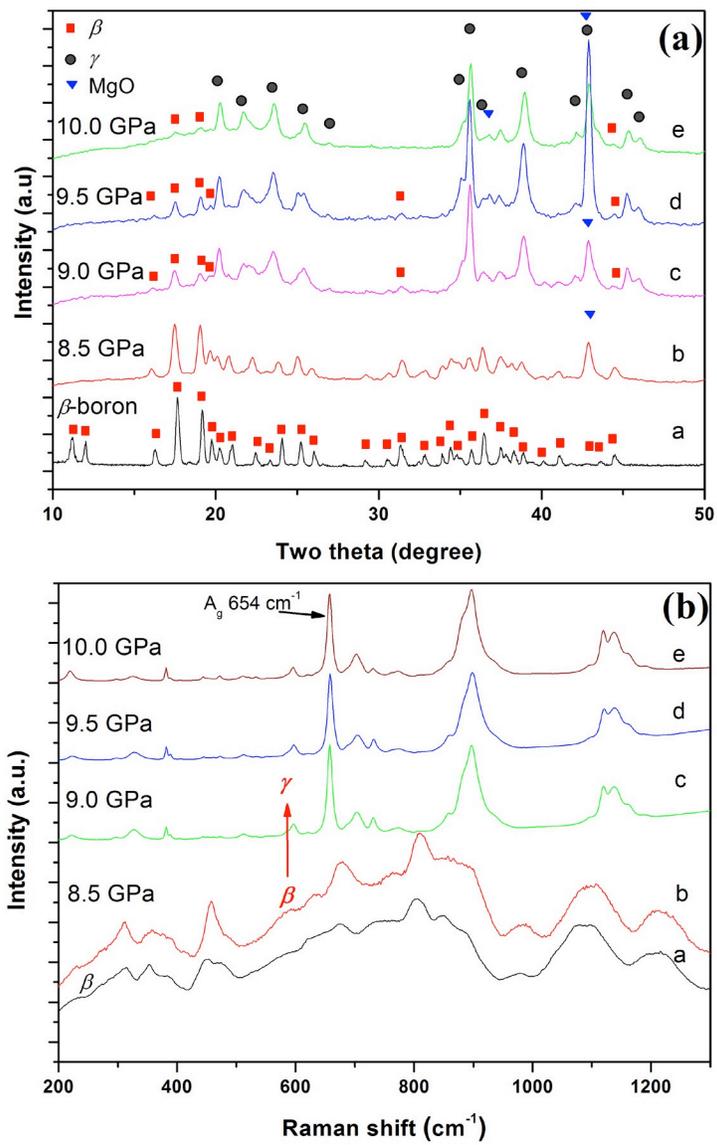

Fig.3

Fig.4